\documentclass[11pt, a4paper]{article}
\usepackage{tikz}
\usetikzlibrary{snakes}
\usepackage[utf8]{inputenc} 
\usepackage{textcomp} 
\usepackage{flafter}  

\usepackage{amsmath,amssymb}  
\usepackage{bm}  
\usepackage{authblk}
\usepackage{memhfixc}  
\usepackage{listings}
\newcommand{\be}{\begin{equation}}
\newcommand{\ee}{\end{equation}}

\def\Xint#1{\mathchoice
{\XXint\displaystyle\textstyle{#1}}%
{\XXint\textstyle\scriptstyle{#1}}%
{\XXint\scriptstyle\scriptscriptstyle{#1}}%
{\XXint\scriptscriptstyle\scriptscriptstyle{#1}}%
\!\int}
\def\XXint#1#2#3{{\setbox0=\hbox{$#1{#2#3}{\int}$}
\vcenter{\hbox{$#2#3$}}\kern-.5\wd0}}

\def\dashint{\Xint-}

\numberwithin{equation}{section}
\begin{document}
\title{Survival probabilities and rates derived from an exact Green's function of the reversible diffusion-influenced reaction for an isolated pair in 2D}
\author{Thorsten Pr\"ustel} 
\author{Martin Meier-Schellersheim} 
\affil{Laboratory of Systems Biology\\National Institute of Allergy and Infectious Diseases\\National Institutes of Health}
\maketitle
\let\oldthefootnote\thefootnote 
\renewcommand{\thefootnote}{\fnsymbol{footnote}} 
\footnotetext[1]{Email: prustelt@niaid.nih.gov, mms@niaid.nih.gov} 
\let\thefootnote\oldthefootnote 
\abstract
{
Recently, an exact Green's function of the diffusion equation for a pair of spherical interacting particles in two dimensions subject to a backreaction boundary condition was derived.
Here, we use the obtained Green's function to calculate exact expressions for the survival probability, the time-dependent reaction rate coefficient for the initially unbound pair and the survival probability of the bound state in the time domain. Moreover, we derive an exact expression for the off-rate.
}
\section{Introduction}
In \cite{TP_MMS_GF:2011}, an exact Green's function (GF) $g(r, t\vert r_{0}) $ of the two dimensional (2D) diffusion equation for a pair of spherical interacting particles was derived. The GF satisfies the backreaction boundary condition \cite{Agmon:1984, Agmon:1990p10, kimShin:1999, TP_MMS_GF:2011} 
\begin{eqnarray}\label{BC}
2\pi a D \frac{\partial}{\partial r}g(r, t\vert r_{0})\vert_{r=a} = \kappa_{a}g(a, t\vert r_{0}) - \kappa_{d}[1-S(t\vert r_{0})].
\end{eqnarray} 
Here, $D$ is the sum of the particles' diffusion constants, $r$ is the inter-particle distance and $a$ is the encounter radius. Furthermore, $\kappa_{a}$ and $\kappa_{d}$ denote the intrinsic association and dissociation rate constants, respectively. 
Finally, $S(t\vert r_{0})$ refers to the probability that a pair of molecules with initial distance $r_{0}$ survives until time $t$, 
\begin{eqnarray}
S(t\vert r_{0})  & = & 2\pi\int^{\infty}_{a} g(r, t\vert r_{0}) r dr \label{defS}\\
& = & 1 - 2\pi a D \int^{t}_{0} \frac{\partial}{\partial r}g(r, t' \vert r_{0})\vert_{r = a}dt'. \label{defSII} 
\end{eqnarray}
It was shown that the associated GF takes the form \cite{TP_MMS_GF:2011}
\begin{equation}\label{exactGF}
g(r, t\vert r_{0} )=\frac{1}{2\pi}\int^{\infty}_{0}e^{-Dx^{2}t}T(x, r)T(x, r_{0})\,x\,dx,
\end{equation}
where we introduced the funtions
\begin{eqnarray}\label{T}
T(x, r) = \frac{J_{0}(rx)\beta(x) - Y_{0}(rx)\alpha(x)}{[\alpha(x)^{2}+\beta(x)^{2}]^{1/2}},
\end{eqnarray}
and
\begin{eqnarray}
\alpha(x) &:=& ( x^{2} - \kappa_{D})J_{1}(xa) + hxJ_{0}(xa), \\
\beta(x) &:=& ( x^{2} - \kappa_{D})Y_{1}(xa) + hxY_{0}(xa). 
\end{eqnarray}
$J_{0}, J_{1}, Y_{0}, Y_{1}$ denote the Bessel functions of first and second kind and of zeroth and first order, respectively \cite{abramowitz1964handbook}. Furthermore, by definition we have
\begin{eqnarray}
h := \frac{\kappa_{a}}{2\pi a D} \\
\kappa_{D} := \frac{\kappa_{d}}{D}.
\end{eqnarray}

Knowing the GF allows us to derive further important quantities, notably, the survival probability. For the following calculations, it turns out to be convenient to introduce the function
\begin{eqnarray}\label{P}
P(x, r) := -\frac{1}{x}\frac{\partial}{\partial r}T(x, r) =  \frac{J_{1}(xr)\beta(x) - Y_{1}(xr)\alpha(x)}{[\alpha(x)^{2}+\beta(x)^{2}]^{1/2}},
\end{eqnarray}
where we used \cite{abramowitz1964handbook}
\begin{eqnarray}\label{J}
J^{\prime}_{0}(x)&:=&\frac{d}{dx}J_{0}(x) = - J_{1}(x) \\
\label{Y}
 Y^{\prime}_{0}(x)&:=&\frac{d}{dx}Y_{0}(x) = - Y_{1}(x).
\end{eqnarray} 
The survival probability may be calculated according to \eqref{defS} or, alternatively, using \eqref{defSII}.  Either way, we find
\begin{equation}\label{revSurv}
S(t\vert r_{0}) = 1 - a \int^{\infty}_{0} e^{-Dt x^{2}} P(x, a) T(x, r_{0})dx.
\end{equation}
To obtain \eqref{revSurv} via \eqref{defSII} we used the integral identity
\be\label{int0} 
\int^{\infty}_{0}P(x, a)T(x, r_{0})dx = 0.
\ee 
\eqref{int0} can be derived by substituting the GF \eqref{exactGF} into the boundary condition \eqref{BC}. Then, using \eqref{P}, \eqref{J} and \eqref{Y}, as well as \cite{abramowitz1964handbook} 
\be\label{comm}
Y^{\prime}_{0}(x)J_{0}(x) - J^{\prime}_{0}(x)Y_{0}(x) = \frac{2}{\pi x},
\ee
we are directly led  to \eqref{int0}.
From \eqref{revSurv} we can easily conclude that the reversible survival probability approaches unity for large times
\begin{equation}
\lim_{t\rightarrow \infty} S(t\vert r_{0}) = 1,
\end{equation}
implying that in 2D the ultimate fate of an isolated pair for the reversible reaction is always dissociation, as in the 1D and 3D case \cite{Agmon:1984, Agmon:1990p10, kimShin:1999}.

Turning now our attention to the initially bound pair, we use the notation $\ast$ to indicate the bound state and we let $g(r, t\vert \ast)$ denote the Green's function describing an initially bound pair that is found separated 
by a distance $r$ at  a later time $t$, cp \cite{Agmon:1990p10, kimShin:1999}. Knowledge of this GF is desirable, because it enables us to compute the average lifetime of the bound state and, hence, the off-rate, as we shall see later. 

We note that $g(r, t\vert \ast)$ is related to the GF  $g(\ast, t\vert r_{0})$, which describes an initially unbound pair that is bound at a later time $t$, by the detailed balance condition \cite{Agmon:1990p10, kimShin:1999}
\begin{equation}\label{detailedBalance}
\kappa_{d} g(*, t\vert r) =  \kappa_{a} g(r, t\vert *).
\end{equation}
Furthermore, we have
\begin{equation}\label{BoundpSurvivalUnity}
g(*, t\vert r_{0}) + S(t\vert r_{0})= 1 .
\end{equation}  
\eqref{BoundpSurvivalUnity} makes it evident that, technically speaking, the GF $g(*, t\vert r_{0})$ represents a probability rather than a probability density.

From \eqref{revSurv}, \eqref{detailedBalance} and \eqref{BoundpSurvivalUnity} it follows that
\begin{equation}
g(r, t\vert *) = \frac{\kappa_{d}}{\kappa_{a}} a \int^{\infty}_{0}e^{-Dt x^{2}} P(x, a) T(x, r)dx.
\end{equation}
Upon direct integration we arrive at the probability $S(t\vert *)$ that an initially bound pair is unbound at time $t > 0$ \cite{Agmon:1990p10, kimShin:1999}  
\begin{equation}\label{S*}
S(t\vert *) =2\pi\int^{\infty}_{a}r\,g(r, t\vert *)dr = 1  - 2\pi\frac{\kappa_{d}}{\kappa_{a}} a^{2} \int^{\infty}_{0}e^{-Dt x^{2}} P^{2}(x, a) \frac{1}{x}dx.
\end{equation}
Note that $S(0 \vert *) = 0$, due to \eqref{int2}. 

In \cite{Agmon:1990p10} it was demonstrated that $S(t\vert *)$ relates to $S(t\vert r_{0})$ in the time domain in the following way:
\be\label{S*S}
S(t\vert *) = \kappa_{d}\int^{t}_{0} e^{-\kappa_{d}t^{\prime}}S(t-t^{\prime}\vert a)dt^{\prime}.
\ee
Using \eqref{T}, \eqref{P}, \eqref{revSurv}, \eqref{comm},  and \eqref{S*}, we can now explicitly show that \eqref{S*S} is indeed satisfied in 2D.  

\section{Rates}
In the Smoluchowski approach, steady-state rate constants can be calculated in two related ways. One method employs the irreversible time-dependent solution of the diffusion equation to calculate the time-dependent rate coefficient $k_{\text{irr}}(t)$. The steady-state rate constant $k^{\text{on}}_{\text{irr}}$ is then given by the long time limit of $k_{\text{irr}}(t)$
\be
k^{\text{on}}_{\text{irr}}:= \lim_{t\rightarrow\infty}k_{\text{irr}}(t)
\ee
While in 3D this procedure yields  the classic result $k^{\text{on}}_{\text{irr}} = 4\pi a D$ (for purely absorbing boundary conditions), in 2D one obtains  the result that $k^{\text{on}}_{\text{irr}}$ actually vanishes.
This issue is also reflected in the second approach, which is based on steady-state solutions of the diffusion equation. In 2D, however, the steady-state solutions are logarithmic, and hence, no boundary condition can be imposed at infinity. As a remedy, several procedures have been proposed \cite{rich1968structural, berg1977physics, keizer1987diffusion, lauffenburger1996receptors}. All of them give rise to another length scale $b > a$ in addition to the encounter radius. Furthermore, they lead, up to a constant, to the same expression for  the on-rate constant 
\be\label{konApprox}
\frac{1}{k_{\text{on}}} = \frac{1}{\kappa_{a}} + \frac{\ln(b/a) + \delta}{2\pi D}.
\ee 
Using $K_{\text{eq}}:=\frac{\kappa_{a}}{\kappa_{d}}, k^{-1}_{\text{off}} = K_{\text{eq}} k^{-1}_{\text{on}}$ one arrives at an expression for the off-rate      
\be\label{koffApprox}
\frac{1}{k_{\text{off}}} = \frac{1}{\kappa_{d}} + K_{\text{eq}}\frac{\ln(b/a) + \delta}{2\pi D}.
\ee 
Here $\delta$ denotes the above mentioned constant that assumes a varying value depending on the chosen method.

In the following we will use the reversible GF \eqref{exactGF} to address these issues.  
The time-dependent reaction rate coefficient can be obtained from the survival probability $S(t\vert r_{0})$ \cite{Agmon:1990p10, kimShin:1999} 
\begin{equation}\label{k}
k(t) = 2\pi a D \frac{\partial}{\partial r_{0}} S(t\vert r_{0})\vert_{r_{0} = a}.
\end{equation}
Using \eqref{revSurv}, we obtain the exact expression in the time domain 
\begin{equation}\label{timeRate}
k(t) = 2\pi a^{2} D \int^{\infty}_{0} e^{-Dt x^{2}} P^{2}(x, a) x dx.
\end{equation}
It follows immediately from \eqref{timeRate} that the long time limit of $k(t)$ vanishes.

As shown in the appendix, the integral $\int^{\infty}_{0}P^{2}(x, a) \frac{dx}{x}$ can be computed analytically and we find
\begin{equation}\label{integralIdentity}
\int^{\infty}_{0}P^{2}(x, a) \frac{dx}{x} = \frac{\kappa_{a}}{\kappa_{d} 2\pi  a^{2}}.
\end{equation}
With the help of this equation and \eqref{timeRate} we recover the correct expression for the equilibrium constant $ K_{\text{eq}}$
\begin{equation}\label{Keq}
\int^{\infty}_{0}k(t) dt = \frac{\kappa_{a}}{\kappa_{d}} =: K_{\text{eq}}.
\end{equation}

Alternatively, identity \eqref{Keq} may be deduced in the following way. 
According to reference \cite{Agmon:1990p10} the time-dependent reaction rate can also be calculated by using the survival probability $S(t\vert *)$, cp \eqref{k}
\be\label{kS*}
k(t) = K_{\text{eq}}\frac{\partial S(t\vert *)}{\partial t}.
\ee
Now, given \eqref{revSurv} and \eqref{S*}, one can explicitly show that \eqref{k} and \eqref{kS*} yield the same result. Then, taking into account that $S(\infty\vert *) = 1$ and  $S(0\vert *) = 0$, \eqref{Keq} is 
immediately implied by \eqref{kS*}.

The steady-state dissociation rate constant, or off-rate, is defined by \cite{Agmon:1990p10, kimShin:1999}
\be
k_{\text{off}} = \tau^{-1},
\ee  
where $\tau$ refers to the average lifetime of the bound state.  $\tau$ can be calculated according to \cite{Agmon:1990p10, kimShin:1999}
\be
\tau = \int^{\infty}_{0}[1 - S(t\vert *)] dt . 
\ee

Using \eqref{S*} and after transforming from the dummy integration variable $x$ to the dimensionless integration variable $ \xi := ax$ we arrive at
\be\label{avLifeTime}
\tau = 2\pi\frac{\kappa_{d}}{\kappa_{a}}\frac{a^{4}}{D} \int^{\infty}_{0}\frac{f(\xi)}{\xi} d\xi,
\ee
where we have introduced the function
\be
f(\xi):=\frac{P^{2}(\xi, 1)}{\xi^{2}}
\ee
for notational convenience. We note that the integrand in  \eqref{avLifeTime} depends on the dimensionless constants $\tilde h = ha, \tilde \kappa_{D}:= \kappa_{D} a^{2}$ only, due to the transformation $x\rightarrow \xi = xa$.
Since 
\be
\lim_{\xi\rightarrow 0}f(\xi) = \frac{\tilde h^{2}}{\tilde \kappa^{2}_{D}}\neq 0,
\ee
it follows that the integrand is singular at the lower endpoint $\xi=0$ of the integration interval, which implies that the integral does not exist.  However, one can still assign a well-defined value to integrals of that type upon regularizing them in the sense of Hadamard's one-sided finite-part integrals \cite{kythe2005handbook}
\begin{eqnarray}\label{hadamard}
\dashint^{\infty}_{0} \frac{f(\xi)}{\xi}d\xi:= \int^{\infty}_{c}\frac{f(\xi)}{\xi}d\xi 
+\int^{c}_{0} \frac{f(\xi)-f(0)}{\xi}d\xi + f(0)\dashint^{c}_{0}\frac{d\xi}{\xi}.
\end{eqnarray}
By definition, the following finite part integral yields \cite{kythe2005handbook}
\be 
\dashint^{c}_{0}\frac{d\xi}{\xi}:= \ln c
\ee
We would like to point out that one has to split the integral at $\infty > \xi = c > 0$ in \eqref{hadamard}, because otherwise the regularization would introduce another singularity due to the occurrence of $\ln\infty$ terms.
It is instructive to compare the resulting expression for $\tau$ with \eqref{koffApprox}.
Clearly, the third term on the rhs of (\ref{hadamard}) gives rise to precisely the logarithmic contribution if one set $c=b/a$. Moreover, one can show numerically that the other two integrals in \eqref{hadamard} yield
\be
\int^{\infty}_{c}\frac{f(\xi)}{\xi}d\xi + \int^{c}_{0}\frac{f(\xi)-f(0)}{\xi}d\xi = \frac{\tilde h}{\tilde \kappa^{2}_{D}} + C(c) \frac{\tilde h^{2}}{\tilde \kappa^{2}_{D}},
\ee
where $C(c)$ depends on the choice of $c$ only. Taken together this means that one seemingly recovers the result (\ref{koffApprox}). However, it is easily demonstrated that the regularized integral is actually independent of the choice of $c$. Thus, in particular, one can choose $c =1$, which shows that the logarithmic contribution actually vanishes and one obtains
\be
\tau = \frac{1}{k_{\text{off}}} = \frac{1}{\kappa_{d}} + K_{\text{eq}}\frac{C}{2\pi D},
\ee 
where $C:=C(1)\approx 0.11593....$. We find the result that the off-rate does not depend on the encounter radius.
\section{Appendix}
In this appendix we will compute the integrals that are needed in the main text.

The GF has to satisfy the initial condition \cite{TP_MMS_GF:2011}
\begin{equation}
g(r, t=0 \vert r_{0}) = \frac{\delta(r-r_{0})}{2\pi r_{0}},
\end{equation}
i.e. in the case considered here we have
\begin{equation}\label{init}
\frac{1}{2\pi}\int^{\infty}_{0} T(x, r) T(x, r_{0}) x dx = \frac{\delta(r - r_{0})}{2\pi r_{0}}.
\end{equation}
From \eqref{init} it follows by direct integration over $\int^{r}_{a}dr' r'$
\begin{equation}\label{ID3}
\int^{\infty}_{0}P(x, r) T(x, r_{0})dx = \biggl\{\begin{array}{lr}
0, &\text{$\quad   r < r_{0}$},\\
r^{-1}, &\text{$\quad   r > r_{0}$,}
\end{array}
\end{equation}
where we have used \eqref{int0}. Furthermore, to perform the integral we used \cite{abramowitz1964handbook}
\be
\frac{d}{dx}[x^{\nu}J_{\nu}(x)] = x^{\nu} J_{\nu-1}
\ee
and
\be
\frac{d}{dx}[x^{\nu}Y_{\nu}(x)] = x^{\nu} Y_{\nu-1}.
\ee
Next, we apply the derivative $\frac{\partial}{\partial r_{0}}$ to the boundary condition \eqref{BC} and subsequently integrate from $0$ to $t$. Taking the limit $t\rightarrow\infty$ leads to 
\begin{eqnarray}
\int^{\infty}_{0}P(x, a) P(x, r_{0}) x dx = -\frac{\kappa_{a}}{2\pi a D} \int^{\infty}_{0}T(x, a) P(x, r_{0})dx + \nonumber\\\frac{\kappa_{d}}{D}\int^{\infty}_{0}P(x, a) P(x, r_{0})\frac{dx}{x}.\qquad\qquad\qquad
\end{eqnarray}
The integral on the lhs has to vanish, as can be seen by taking into account (\ref{int0}) and 
\begin{equation}
\int^{\infty}_{0}P(x, a) P(x, r_{0}) x dx = - \frac{\partial}{\partial r_{0}} \int^{\infty}_{0}P(x, a) T(x, r_{0})dx.
\end{equation}
 Furthermore, we just calculated the first integral on the rhs, cp Eq.~(\ref{ID3}). Hence, we arrive at
\begin{equation}\label{int2}
\int^{\infty}_{0}P(x, a) P(x, r_{0})\frac{dx}{x} = \frac{\kappa_{a}}{\kappa_{d}}\frac{1}{2\pi a r_{0}}.
\end{equation} 
\subsection*{Acknowledgments}
This research was supported by the Intramural Research Program of the NIH, National Institute of Allergy and Infectious Diseases. 

We would like to thank Bastian R. Angermann and Frederick Klauschen for helpful and stimulating discussions.

\bibliographystyle{utphys} 
\bibliography{Off_Rate2D} 

\providecommand{\href}[2]{#2}\begingroup\raggedright\begin{thebibliography}{10}

\bibitem{TP_MMS_GF:2011}
T.~Pr{\"u}stel and M.~Meier-Schellersheim
  \href{http://arxiv.org/abs/1109.4465}{{\ttfamily arXiv:1109.4465 [q-bio]}}.

\bibitem{Agmon:1984}
N.~Agmon {\em J. Chem. Phys.} {\bfseries 81} (1984) 2811.

\bibitem{Agmon:1990p10}
N.~Agmon and A.~Szabo {\em J. Chem. Phys.} {\bfseries 92} (1990) 5270.

\bibitem{kimShin:1999}
H.~Kim and K.~Shin {\em Phys. Rev. Lett.} {\bfseries 82} (1999) 1578.

\bibitem{abramowitz1964handbook}
M.~Abramowitz and I.~Stegun, {\em Handbook of Mathematical Functions with
  Formulas, Graphs, and Mathematical Tables}.
\newblock Dover, New York, 1965.

\bibitem{rich1968structural}
G.~Adam and M.~Delbr{\"u}ck, {\em Structural chemistry and molecular biology,
  {eds. Rich, A. and Davidson, N.R.}}
\newblock W. H. Freeman, 1968.

\bibitem{berg1977physics}
H.~Berg and E.~Purcell {\em Biophys. J.} {\bfseries 20} (1977) 193.

\bibitem{keizer1987diffusion}
J.~Keizer {\em Chem. Rev.} {\bfseries 87} (1987) 167.

\bibitem{lauffenburger1996receptors}
D.~Lauffenburger and J.~Linderman, {\em Receptors: models for binding,
  trafficking, and signaling}.
\newblock Oxford University Press, 1993.

\bibitem{kythe2005handbook}
P.~Kythe and M.~Sch{\"a}ferkotter, {\em Handbook of computational methods for
  integration, v. 1}.
\newblock Chapman \& Hall/CRC, 2005.

\end{thebibliography}\endgroup

\end{document}